\documentclass[conference]{IEEEtran}

\usepackage{cite}
\usepackage{dsfont}
\usepackage{amsmath,amsthm,amssymb,amsfonts,bm,xparse}
\newtheorem{theorem}{Theorem}

\newtheorem{lemma}{Lemma}
\newtheorem{problem}{Problem}
\newtheorem{proposition}{Proposition}
\newtheorem{game}{Game}
\newtheorem{question}{Question}
\newtheorem{corollary}{Corollary}
\newtheorem{assumption}{Assumption}

\usepackage{stfloats}

\usepackage{cases}
\newtheorem{defn}{Definition}
\usepackage{algorithmic}
\usepackage{graphicx}
\usepackage{epstopdf}
\usepackage{textcomp}
\usepackage{xcolor}
\usepackage{url}
\usepackage{enumitem}
\usepackage{url}
\usepackage{bbm}
\usepackage{tcolorbox}
\setlist[itemize]{leftmargin=*}
\setlist[enumerate]{leftmargin=*}
\usepackage{multirow}
 
\usepackage{subfigure}
\usepackage[linesnumbered, ruled]{algorithm2e}
\def\BibTeX{{\rm B\kern-.05em{\sc i\kern-.025em b}\kern-.08em
    T\kern-.1667em\lower.7ex\hbox{E}\kern-.125emX}}
\begin{document}
\title{Strategic Data Revocation in Federated Unlearning 
}
\IEEEoverridecommandlockouts
\author{\IEEEauthorblockN{Ningning Ding}
\IEEEauthorblockA{Northwestern University, USA \\
ningning.ding@northwestern.edu}
\and
\IEEEauthorblockN{Ermin Wei}
\IEEEauthorblockA{Northwestern University, USA\\
ermin.wei@northwestern.edu}
\and
\IEEEauthorblockN{Randall Berry}
\IEEEauthorblockA{Northwestern University, USA\\
rberry@northwestern.edu}
\thanks{This work is supported by NSF  ECCS-2030251 and NSF ECCS-2216970.}
}

\maketitle
\begin{abstract} 
By allowing users to erase their data's impact on federated learning models, federated unlearning protects users' right to be forgotten and data privacy. 
Despite a burgeoning body of research on federated unlearning's technical feasibility, there is a paucity of literature investigating the considerations behind users' requests for data revocation. 
This paper proposes a non-cooperative game
framework to study users' data revocation strategies in federated unlearning. We prove the existence of a Nash equilibrium. However, users' best response strategies are coupled via model performance and unlearning costs, which makes the equilibrium computation challenging. We obtain the Nash equilibrium by establishing its equivalence with a much simpler auxiliary optimization problem. 
We also summarize users' multi-dimensional attributes into a single-dimensional metric  and derive the closed-form characterization of an equilibrium, when users' unlearning costs are negligible. 
Moreover, we compare the cases of allowing and forbidding partial data revocation in federated unlearning. Interestingly, the results  reveal that allowing partial revocation does not necessarily increase users' data contributions or payoffs due to the game structure. 
Additionally, we demonstrate that positive externalities may exist between users'  data revocation decisions when users incur unlearning costs, while this is not the case when their unlearning  costs are negligible.
\end{abstract}
\section{Introduction}
\subsection{Background and Motivations}
\emph{Federated unlearning} is an emerging technique designed to erase the influence of certain users' data  from  a  trained federated learning model. This is motivated by  the success of {federated learning} over large numbers of edge users and the recent regulations guaranteeing data owners' rights to be forgotten. 

Federated learning is a distributed machine learning paradigm, where many users collaboratively train a shared learning model under a server’s coordination. 
Although federated learning helps protect privacy by allowing distributed training  without sharing users' raw data, the trained model can still leak users'  information (e.g., through a backdoor attack) \cite{wang2020attack,lyu2020threats}.
This motivates recent regulations to protect users' data privacy and their rights to revoke their data from a trained model.  Examples include the European Union General Data Protection Regulation (GDPR) \cite{voigt2017eu} and the California Consumer Privacy Act (CCPA) \cite{harding2019understanding}. A naive strategy to accomplish this is to retrain the model from scratch using the remaining users' data, which is time-consuming and computationally expensive. Therefore, federated unlearning   methods  focus  on removing the influence of revoked data without retraining the whole model from scratch. Typically, this process needs the staying
users   to perform additional calculations based on the learned model and remaining data to  obtain an unlearned model \cite{liu2020learn}.

Given the right to be forgotten, users may seek to revoke their data from the learned model for a variety of reasons; a few examples follow. First,  if users perceive their benefits from the learned model as insufficient to offset  their costs, they may revoke their data. Second, users typically  lack full knowledge of the associated benefits (e.g., model performance) and costs (e.g.,  privacy costs tied to data uniqueness) until they participate in  federated learning  \cite{jiang2019improving}. 
Consequently, if the actual outcomes deviate from their initial estimations, users may request  their data to be unlearned.
Third, users can mitigate or even eliminate certain costs (e.g.,  data privacy costs) by revoking their data  after participating in federated learning  \cite{gao2022verifi}.

It is challenging for users to optimally decide  whether to revoke data   and how much data to revoke. Users need to make complex assessments of the benefits and costs associated with their actions. These considerations include the local performance of the trained model, their potential privacy leakage from participation, and federated unlearning costs if they stay in the system (i.e., additional training for obtaining an unlearned model).
Moreover, users' strategies for data revocation indirectly impact each other's payoffs. For example, each user's  data revocation action will affect the   performance of the unlearned global model and   the required efforts for federated unlearning  (e.g., required training rounds for users). It is hard for each user  to precisely anticipate such impacts and make optimal decisions, especially given a large number of heterogeneous users with multi-dimensional attributes on data and payoffs. 
This motivates our first key question:
\begin{question}
	What are heterogeneous users' optimal data revocation strategies in federated unlearning?
\end{question}

Furthermore, existing federated unlearning literature (e.g., \cite{liu2021federaser,liu2020learn,ding2023inc}) usually operated under the assumption that users either completely leave the system (i.e., revoke all their data)  or remain entirely committed (i.e., do not revoke any data). Partial data revocation remains relatively uncharted territory. However, this may harbor  the potential to increase  the total  data size remaining in the system and users' payoffs, compared to the scenario where   partial data revocation is forbidden. This lead to the second key question in this paper:
\begin{question}
	Compared with forbidding partial data revocation, is allowing partial  revocation more beneficial in terms of    users' payoffs and total remaining  data size?
\end{question}

\subsection{Contributions}
We summarize our key contributions below.
\begin{itemize}
	\item \emph{Data revocation in federated unlearning:} 	We propose a game-theoretic model to characterize the interaction among users  in federated unlearning. 
	To the best of our knowledge, this is the first analytical work to  study users' strategic data revocation in federated unlearning, with partial data revocation allowed.
	
	\item \emph{Users'  data revocation strategies at equilibrium:} Given the complicated mutual influence of users' decisions through model performance and unlearning costs, the problem is challenging to analyze (e.g., with non-convex, non-monotonic, and piece-wise best responses). 
	We  calculate the equilibrium through   a problem transformation that applies to general payoff functions. 
	Moreover, when users' unlearning costs are negligible, we summarize users' multi-dimensional attributes into	a single-dimensional metric that enables us to give a closed-form  equilibrium. 

	\item \emph{Comparison between allowing and forbidding partial data revocation:}
	We investigate whether allowing users' partial revocation is better than when it is forbidden. The results counter-intuitively show that  allowing partial revocation may not increase users'  remaining data amount or   payoffs   because of the strategic interactions among users. 
	
	\item \emph{Insights:} 	
	Our results reveal   trade-offs among model performance,  data privacy concerns, and federated  unlearning costs, highlighting how these are influenced by factors such as model dependency, data size, and data uniqueness. 
	We also show that  users may  follow the others' data revocation decisions (i.e., a positive externality) when they incur unlearning costs but will not (i.e., a negative externality) when unlearning costs are negligible. 
	 This is because the model performance  increases in users'   remaining data sizes while unlearning costs increase in users' revoked data sizes.
	
\end{itemize}

\subsection{Related Work}
\emph{Machine unlearning} was first introduced in 2015 \cite{cao2015towards}. It focuses on   how to efficiently erase the effects of certain data on a \emph{centrally} trained  machine learning model. 
The current machine unlearning methods often involve properly  modifying training data or altering the trained model parameters to reduce the model's reliance on the deleted information (e.g.,  \cite{nguyen2022survey}). 
However, the server in a federated network has no access to the local datasets of a large number of users, which makes data modification and manipulation impossible. Moreover, users in a federated system contribute to the final model through iterative training processes, which makes it not straightforward to partition data impact and alter model parameters.

%

Some pioneering  works  have recently proposed   federated unlearning  mechanisms using methods such as  gradient subtraction (e.g., \cite{liu2021federaser,liu2020learn}),   gradient scaling (e.g., \cite{gao2022verifi,liu2022right}), or knowledge distillation  (e.g., \cite{wu2022federated}).  These papers did not study which user(s) will revoke data and how much data will be revoked. This is of great importance in understanding users' behaviors  and developing related mechanisms or regulations.  To fill this gap, we focus on users'   data revocation strategies in this paper.

Furthermore, there is a wide spectrum of literature on studying (or incentivizing) users'   strategic data contributions in federated learning, utilizing game theory or other  economic methodologies (e.g., \cite{zhan2020learning,ding2020optimal,zhang2021faithful,wang2022socially,zhang2022enabling}). However, these studies did not incorporate the unique aspects of federated unlearning (e.g., unlearning costs), so it is hard to apply their results to our context.
Ding et al. in \cite{ding2023inc} proposed an incentive mechanism for federated unlearning, aimed at retaining strategic users intending to leave the system. Nevertheless, their assumption of  all-or-nothing user engagement fails to account for scenarios where users may wish to partially revoke their data. They  also assumed that users do not care about the global model's performance. 
To the best of our knowledge, this paper is the first to study  users' strategic data revocation   in federated unlearning,  considering a more general scenario with users' partial data revocation and  interests in the model performance.


The rest of the paper is organized as follows. We first introduce the system model   in Section \ref{model}. In Section \ref{optimal}, we analyze users'  equilibrium strategies on data revocation. To give more insights, we study a special scenario in Section \ref{negaligible}, where users' unlearning costs are negligible. In Section \ref{compareaf}, we investigate whether allowing partial revocation is beneficial compared with when it is forbidden.  We present simulation results in Section \ref{simulation} and conclude this paper in Section \ref{conclusion}.

\section{System Model}
\label{model}
We consider a setting in which a set of users have already participated in federated learning using their local data and are faced with a federated unlearning decision. 
The heterogeneous users now simultaneously decide the amount of data  to revoke from the learned model  and collaboratively accomplish the unlearning process. We  first introduce the objectives and process of federated learning and unlearning, and then we specify users' strategies  and   payoffs, respectively. Finally, we frame a non-cooperative game model, highlighting the users' participation within the system.

\subsection{Federated Learning and Federated Unlearning}
\subsubsection{Federated Learning}
\label{learningo}
Consider a set $\mathcal{I}=\{1,2,...,I\}$ of users.
The purpose of federated learning is to compute a model parameter $w$ by using all users' local data. The optimal model parameter $w^*$  minimizes  the global loss function, which is an  average of all users' loss functions $\{F_{i}(w)\}_{i\in \mathcal{I}}$ \cite{karimireddy2020scaffold,pathak2020fedsplit}:
\begin{equation}
\label{weight}
w^*=\arg\min _{w}\frac{1}{I}\sum_{i\in \mathcal{I}} F_{i}(w).
\end{equation}
During the federated learning process, each user computes a local model update  based on local data and sends the model update  to a central server.  The server aggregates the local updates and sends the global shared model back to the users for the next round of training. The process is then repeated until the global model converges \cite{mcmahan2016communication}.

%


\subsubsection{Federated Unlearning}
\label{unlearningo}
A federated learning process   maps  users' data  into a model space, while a federated unlearning process  maps a  learned model,  users' data set, and the data set that is required to be forgotten into  an unlearned model space. The goal of federated unlearning is to  make the unlearned model have the same distribution   as a retrained model (i.e., retrained from scratch using the remaining data).\footnote{The distribution is due to the randomness in the training process (e.g. randomly sampled data and random ordering of batches).}


A natural method for federated unlearning  is to let the staying users (excluding leaving users) use the remaining data to continue training  from the learned model $w^*$. The training continues until it converges to  a new optimal model parameter $\tilde{w}^*$, which minimizes the global loss function of staying users: 
\begin{equation}
\label{weight1}
\tilde{w}^*=\arg\min _{w}\frac{1}{I-I_{leave}}\sum_{i\in \mathcal{I}\backslash\mathcal{I}_{leave}} \tilde{F}_{i}(w),
\end{equation}
where   $\mathcal{I}_{leave}$ is the set of  users who leave the system through federated unlearning. This method is typically more efficient than training from scratch, as the minimum point may not change much after some users leave.
As in \eqref{weight1}, the unlearned model greatly depends on users' data revocation strategies.

\subsection{Users' Strategies}
We use  $d_i^{\max}>0$ to denote the  size of  data that a user $i\in \mathcal{I}$ has previously utilized in training a federated learning model. Thus, $d_i^{\max}$ is the maximum data size that user $i$ can revoke through the unlearning phase. 

Each user $i$ decides the   data size $d_i\in [0,d_i^{\max}]$ to \emph{remain} in the system after revocation.  In other words, the data size that user $i$  revokes  through federated unlearning is $d_i^{\max}-d_i$.\footnote{This paper focuses on analyzing the case of allowing partial data revocation. We will compare it with the case of forbidding partial revocation (i.e., binary decision $d_i\in \{0,d_i^{\max}\}$) in Section \ref{compareaf}.} We assume that the revoked data is  uniformly sampled at random from $d_i^{max}$, so that the remaining data distribution does not change. Users' different data revocation strategies will inevitably result in diverse payoffs.

\subsection{Users' Payoffs}
Each user $i$'s payoff from data revocation consists of three parts. 
\subsubsection{Global Model's Performance}
To capture the global model's performance on user $i$'s local data, we employ a widely-adopted model that provides a good approximation of the experimental statistics (e.g., \cite{zhan2020learning,zhan2020big,hastie2009elements,lu2009similarity}):
\begin{equation}
\label{e3}
P_i(d_i,\boldsymbol{d_{-i}})=\ln\left(\sum_{j\in\mathcal{I}}d_j+\epsilon_i\right),
\end{equation}
where $\boldsymbol{d_{-i}}\triangleq\{d_j\}_{j\in\mathcal{I},j\neq i}$ denotes the  strategies of other users except for user $i$. 
The logarithmic concave model  captures the diminishing marginal utility of additional data. Intuitively, the model performance increases in the total data size remaining in the system. However, when the remaining data size is already substantial, any further increase in the  data size does not significantly improve the model performance. 
The factor $\epsilon_i$ captures the heterogeneity in model performance   and  bounds   user $i$'s cost when all of the users' data is revoked, i.e. $\sum_{j\in\mathcal{I}}d_j=0$. We refer to $\epsilon_i$ as the \emph{independence index}, where a small value (e.g., $\epsilon_i \rightarrow 0$) implies that the user has a high dependency on the global model. 


%
%
	
\subsubsection{Privacy Cost}
	A user $i$'s perceived data privacy cost increases in its remaining data size and the uniqueness of its data (e.g., \cite{de2013unique,romanini2021privacy}):
	\begin{equation}
 \label{e4}
	C_i^p(d_i)=\xi_id_i\ell_i,
	\end{equation}
	where $\xi_i$ is user $i$'s \emph{marginal privacy cost} and $\ell_i$ is user $i$'s \emph{data uniqueness level}.  The data uniqueness level can be characterized by the computed gradient $\Vert \nabla F_i(w^*) \Vert $, as the gradient reflects the distance of user $i$'s data from the average of other users' distributions. For simplicity, we define $\ell_i=\Vert \nabla F_i(w^*) \Vert $. 

\subsubsection{Unlearning Cost}

  According to 
our previous analysis based on  Scaffold and Fedavg algorithms \cite{ding2023inc},  the number of unlearning communication rounds increases in the uniqueness levels  of the users' revoked data:
	\begin{equation}
	\label{e5} \sum_{j\in\mathcal{I}}\left(1-\frac{d_j}{d_j^{max}}\right)\ell_j^2,
	\end{equation}	
 where $(1-d_j/d_j^{\max})$ represents the proportion of user $j$'s revoked data size to its total data size $d_j^{\max}$. When the distance between the learned and  unlearned models is large (as implied by a 
 high proportion and uniqueness of the revoked data), it necessitates an increased number of communication rounds for effective federated unlearning.
 
	User $i$'s unlearning cost   increases in the unlearning rounds and its remained data size (used in each round's unlearning):\footnote{For tractability, we assume that 
 the influence of each user's revocation decision on the unlearning communication round is negligible, 
due to the significant volume of participating users. As a result, each user dismisses its individual impact on the  unlearning rounds (indicated as $j\neq i$ in \eqref{e6}). This assumption is not restrictive in applications like Gboard (Google's keyboard) which encompasses billions of users.}
	\begin{equation}
 \label{e6}
	\begin{split}
	C_i^u(d_i,\boldsymbol{d_{-i}})=\theta_i d_i	\sum_{j\in\mathcal{I}, j\neq i}\hspace{-1mm}\left(1-\frac{d_j}{d_j^{max}}\right)\ell_j^2,
	\end{split}
	\end{equation}
	where $\theta_i$ is user $i$'s marginal unlearning cost.
	

Combining the three parts in \eqref{e3}, \eqref{e4}, and \eqref{e6}, each user $i$'s payoff  is 
\begin{equation}
\label{payofff}
\begin{split}
&U_i(d_i,\boldsymbol{d_{-i}})=P_i(d_i,\boldsymbol{d_{-i}})-C_i^p(d_i)-C_i^u(d_i,\boldsymbol{d_{-i}}).
\end{split}
\end{equation}

As users' payoffs are affected by each other's strategies, the users are engaged in  a game.

\subsection{Game Formulation}
We formally define the resulting game as follows.
\begin{game}[Users' Data Revocation Game]
	\label{game1}
	This game consists of 
	\begin{itemize}
		\item Players: $I$ users in set $\mathcal{I}$.
		\item Strategy space: each user $i\in \mathcal{I}$ decides the data size to remain in the system $d_{i}\in [0,d_i^{\max}]$. 
		\item Payoff function: each user $i \in \mathcal{I}$ maximizes its payoff $U_i(d_i,\boldsymbol{d_{-i}})$ in \eqref{payofff}.
	\end{itemize}
\end{game}
In this game,  each user needs to  trade off the model performance, privacy costs, and unlearning costs, when considering  how much data to revoke. For instance, the contribution of additional data beyond a certain threshold may not significantly enhance the model's performance but lead to substantial privacy and unlearning costs.

As we will see in the next section, each user's  revocation strategy affects other users through the global model performance and unlearning costs in a complex way.  This makes it challenging to analyze the Nash equilibrium of this game.

\section{Users' Data Revocation Equilibrium}
\label{optimal}
In this section, we calculate the Nash equilibrium of  Game \ref{game1}, which is defined as follows. 
\begin{defn}[Equilibrium of Game \ref{game1}]
\label{def}
	The Nash equilibrium of  Game \ref{game1} is a decision profile $\{d_i^*\}_{i\in \mathcal{I}}$,  such that each user $i\in\mathcal{I}$ achieves its maximum payoff assuming other users are following the equilibrium strategies, i.e., 
	\begin{equation}
	U_i(d_i^*,\boldsymbol{d_{-i}^*})\ge  U_i(d_i,\boldsymbol{d_{-i}^*}), \forall d_{i} \in [0,d_i^{\max}].  
	\end{equation}
\end{defn} 

\subsection{Best Responses}
\label{best}
Based on Definition \ref{def}, we  first specify each user's best response to the other users' revocation decisions in Lemma \ref{br}.
\begin{lemma}
	\label{br}
	The best response of user $i\in \mathcal{I}$ is  
	\begin{equation}
	\label{bere}
	\begin{split}
	d_i^*(\boldsymbol{d_{-i}})=\left[\tilde {d_i}\right]_0^{d_i^{\max}}=\min\left\{d_i^{\max},\max \left\{\tilde{d}_i, 0\right\}\right\},
	\end{split}
	\end{equation}
	where
	\begin{equation}
 \label{10}
\hspace{-1mm}\tilde{d}_i\triangleq \hspace{-1mm}\left(\hspace{-1mm}\xi_i \ell_i+\theta_i \hspace{-1mm}\sum_{j\in\mathcal{I}, j\neq i}\hspace{-1mm}\left(1-\frac{d_j}{d_j^{max}}\right)\ell_j^2\hspace{-1mm}\right)^{-1}\hspace{-2mm} - \hspace{-1mm}\sum_{j\in\mathcal{I}, j\neq i}\hspace{-1mm}d_j - \epsilon_i.
	\end{equation}
\end{lemma}

Lemma \ref{br} shows that users with larger costs (i.e.,  $\xi_i$ and $\theta_i$), data uniqueness  level $\ell_i$, and independency index  $\epsilon_i$  tend to revoke more data (i.e., smaller $d_i^*(\boldsymbol{d_{-i}})$). 
However, users' data revocation decisions $\boldsymbol{d}$ influence each other in a complex way. To provide more insights into this interdependency, we consider two extreme cases of Lemma \ref{br}.
\begin{itemize}
	\item  If all other users do not revoke any data, i.e., $d_j= d_j^{\max}, \forall j\in \mathcal{I}, j\neq i$, then user $i$'s best response is
	\begin{equation}
	\begin{split}
	d_i^*(\boldsymbol{d_{-i}})=\left[\left(\xi_i \ell_i\right)^{-1}\hspace{-1mm} - \hspace{-1.5mm}\sum_{j\in\mathcal{I}, j\neq i}\hspace{-1mm}d_j^{\max} \hspace{-1mm}- \epsilon_i\right]_0^{d_i^{\max}}. 
	\end{split}
	\end{equation}
	When the marginal privacy cost $\xi_i$ and data uniqueness level $\ell_i$ are large,  user $i$ will revoke all its data (i.e., $d_i^*=0$) and be a free rider. However, this may not be an equilibrium, as other users may have the incentive to deviate.
	
	\item If all other users fully revoke data, i.e.,  $d_j= 0, \forall j\in \mathcal{I}, j\neq i$, then user $i$'s best response is
	\begin{equation}
	\begin{split}
	d_i^*(\boldsymbol{d_{-i}})=\left[\left(\xi_i \ell_i+\theta_i \hspace{-1mm}\sum_{j\in\mathcal{I}, j\neq i}\hspace{-1mm}\ell_j^2\hspace{-1mm}\right)^{-1} \hspace{-2mm} - \epsilon_i \right]_0^{d_i^{\max}}. 
	\end{split}
	\end{equation}
	When user $i$ has a high dependency on the global model  (i.e., small $\epsilon_i$), it will achieve the unlearning by  itself.\footnote{Note that when only one user remains within the system,  the consequences of leaving or staying are different. 
 By choosing to stay, the user can benefit from a  warm starting point  provided by the platform,  facilitating the training process. However, in this situation, the user still incurs privacy costs. This is because the platform (or server) still can access the model, thus presenting a potential privacy risk.} Specifically, if user $i$ incurs high costs, it will use partial data; if its costs are small, it will not revoke any data. Conversely,  when user $i$ does not really value the global model (i.e., large $\epsilon_i$), it will also revoke all its data (i.e., $d_i^*=0$) like others. Thus, when users' independence indexes $\{\epsilon_i\}_{i\in\mathcal{I}}$ are sufficiently high, all users revoking all their data is an equilibrium. 
\end{itemize}

Based on Lemma \ref{br}, we can also  analyze the mutual influence of users' data (i.e., the externality) as follows:

\begin{corollary}[Externality]
	\label{externality}
	When $d_i^*(\boldsymbol{d_{-i}}) \in (0,d^{\max})$, if 
	\begin{equation}
	\label{c}
	\theta_i\frac{d_j\ell_j^2}{d_j^{\max}}\ge \xi_i\ell_i+\theta_i\sum_{k\in\mathcal{I}, k\neq i,j}(1-\frac{d_k}{d_k^{\max}})\ell_k^2,
	\end{equation}
	then  we have
	\begin{equation}
	\label{r}
	\frac{\partial d_i^*(\boldsymbol{d_{-i}})}{\partial d_j}\ge 0;
	\end{equation}
	otherwise, we have
	\begin{equation}
	\frac{\partial d_i^*(\boldsymbol{d_{-i}})}{\partial d_j}< 0.
	\end{equation}
\end{corollary}
Corollary \ref{externality}  demonstrates that if user $j$'s data uniqueness level $\ell_j
$ and remaining data size $d_j$ are sufficiently large (as indicated by \eqref{c}), user $i$'s partial contribution will increase in user $j$'s remaining data size  (i.e., \eqref{r}).  This suggests that users may  follow others' data revocation decisions, potentially leading to cascaded revocations among users. We will provide illustrative numerical examples of this phenomenon  in Section \ref{strategy}.

As shown in Lemma \ref{br}, the best responses of users are complex piece-wise functions  that incorporate multi-dimensional parameters for user attributes. It is hard to directly calculate the equilibrium based on these best responses, as there will be many possible non-convex and non-monotonic cases depending on   the parameters' values.  We will illustrate the challenge through simulations in Section \ref{bestr}. Next, we first present the Nash equilibrium of homogeneous users  in Section \ref{homo}  and then for heterogeneous users in Section \ref{hete}.

\subsection{Nash Equilibrium of Homogeneous Dependent Users}
\label{homo}
In the homogeneous dependent case, all $I$ users have the same $\theta$, $\ell$, $\xi$, $\epsilon$, and $d^{\max}$, and they  depend on the global  model, i.e., $\epsilon\rightarrow0$. 

The equilibrium in this case is in Proposition \ref{homone}.
\begin{proposition}
	\label{homone}
	The Nash equilibrium of homogeneous dependent users is 
	\begin{itemize}
		\item If $d^{\max} > \frac{1}{I\xi \ell}$, then $\forall i\in \mathcal{I}$,
		\begin{equation}
  \label{partialdata}
		\begin{split}
		&d_i^*=\frac{1}{2\theta \ell (I-1)}\bigg(d^{\max}(\xi+\theta\ell(I-1))- \\&\hspace{-2mm}\left.\sqrt{d^{\max^2}(\xi+\theta\ell(I-1))^2-4\theta(I-1) d^{\max}/I}\right) \in (0,d^{\max}).
		\end{split}
		\end{equation}
  
 \vspace{-3mm}
		\item If $d^{\max} \le \frac{1}{I\xi \ell}$, then 
		\begin{equation}
  \label{alldata}
		d_i^*=d^{\max}, \forall i\in \mathcal{I}.
		\end{equation}
	\end{itemize}
\end{proposition}
Proposition \ref{homone} shows that if the  data size used in federated learning  $d^{\max}$ and the costs $\xi\ell$ are excessively large, users will partially revoke  their data (as in \eqref{partialdata}). This is because contributing more data will not significantly improve the global model's local performance   but will lead to high costs on privacy and unlearning.  
Conversely, when the  amount of data utilized in federated learning  and the costs are relatively small, users will not revoke any data to ensure the good performance of the global model (i.e., \eqref{alldata}).

\subsection{Nash Equilibrium of Heterogeneous Users}
\label{hete}
In this subsection, we focus on the more complex case with heterogeneous users. We first prove the existence and non-uniqueness of the equilibrium. We next present some special equilibria and finally obtain the equilibrium under general conditions.

\subsubsection{Existence and Non-Uniqueness}
\begin{lemma}
	Pure strategy Nash equilibrium exists  but may not be unique in Game \ref{game1}.
\end{lemma}
Game \ref{game1} is a concave game, so a pure strategy Nash equilibrium exists \cite{rosen1965existence}. To illustrate the non-uniqueness, we next present multiple special equilibria that can coexist in Proposition \ref{specialne}. 

\subsubsection{Special Nash Equilibrium}
We first specify the conditions for two special Nash equilibria, where all users have the same strategies: fully revoke or not revoke their data.
\begin{proposition}[Same strategies at equilibrium]
	\label{specialne}
If and only if
	\begin{equation}
	\label{no}
	\epsilon_i\ge \left(\xi_i \ell_i+\theta_i \sum_{j\in\mathcal{I}, j\neq i}\ell_j^2\right)^{-1}, \forall i\in \mathcal{I},
	\end{equation}
there exists an equilibrium where all users fully revoke   their data, i.e., 
	\begin{equation}
 \label{neno}
	d_i^*=0, \forall i\in \mathcal{I}.
	\end{equation}

If and only if
	\begin{equation}
	\label{all}
	\epsilon_i\le \left(\xi_i \ell_i\right)^{-1} - \sum_{j\in\mathcal{I}}d_j^{\max}, \forall i\in \mathcal{I},
	\end{equation} 
there exists an equilibrium where no user   revokes any data: 
	\begin{equation}
 \label{neall}
	d_i^*=d_i^{\max}, \forall i\in \mathcal{I}.
	\end{equation}
\end{proposition}
Note that it is possible for the two conditions \eqref{no} and \eqref{all} to simultaneously hold, so the two equilibria in \eqref{neno} and \eqref{neall} may coexist. This happens when users have highly heterogeneous data (i.e., large data uniqueness $\boldsymbol{\ell}$), large unlearning costs $\boldsymbol{\theta}$, small data sizes $\boldsymbol{d}$, and small privacy concerns $\boldsymbol{\xi}$. 

Proposition \ref{specialne} shows that when the trained model is not very  valuable to users (i.e.,   large independence indexes $\{\epsilon_i\}_{i\in \mathcal{I}}$) and users' costs are significant (i.e., small right-hand side of \eqref{no}), users will revoke all their data. On the other hand, if users really depend on the global model (i.e., small $\{\epsilon_i\}_{i\in \mathcal{I}}$ in \eqref{all}), privacy costs $\xi_i \ell_i$ are small, and there is still room for improving model performance (i.e., contributed data sizes $\sum_{j\in\mathcal{I}}d_j^{\max}$ are small), then all user fully contributing their data without any revocation is an equilibrium. This finding is consistent with the homogeneous dependent case where $\epsilon\rightarrow0$, as described in Proposition \ref{homone}.

There also exists Nash equilibrium where some users fully revoke their data, some users partially revoke   data, and others do not revoke any data. Proposition \ref{different} is an example.
\begin{proposition}[Different strategies at equilibrium]
	\label{different}
	If  conditions \eqref{nor}, \eqref{allr}, and \eqref{partial} on the next page hold,
	\begin{figure*}
			\small
		\begin{equation}
	\label{nor}
	\begin{split}
	\epsilon_i\ge& \left(\xi_i\ell_i+\theta_i\sum_{i'\in\mathcal{I}_1\backslash\{i\}}\ell_{i'}^2+\theta_i\left(1-\left(\left(\xi_k\ell_k+\theta_k\sum_{i'\in \mathcal{I}_1}\ell_{i'}^2\right)^{-1}\hspace{-4mm}-\sum_{j\in\mathcal{I}_2}d_j^{\max}-\epsilon_k\right)/d_k^{\max}\right)\ell_k^2\right)^{-1} \hspace{-4mm} -\hspace{-1mm} \left(\xi_k\ell_k+\theta_k\sum_{i'\in \mathcal{I}_1}\ell_{i'}^2\right)^{-1}\hspace{-4mm}+\epsilon_k,\forall i\in \mathcal{I}_1,
	\end{split}
	\end{equation}
	\vspace{-3mm}
	\begin{equation}
	\label{allr}
	\begin{split}
	\hspace{-5mm}\epsilon_j\le&  \left(\xi_j\ell_j +\theta_j\hspace{-0.5mm}\sum_{i\in\mathcal{I}_1}\ell_{i}^2+\theta_j\hspace{-0.5mm}\left(1-\hspace{-1mm}\left(\hspace{-0.5mm}\left(\xi_k\ell_k+\theta_k\sum_{i\in \mathcal{I}_1}\ell_{i}^2\right)^{-1}\hspace{-2mm}-\hspace{-2mm}\sum_{j'\in\mathcal{I}_2}d_{j'}^{\max}\hspace{-1mm}-\epsilon_k\right)/d_k^{\max}\hspace{-1mm}\right)\hspace{-1mm}\ell_k^2\right)^{-1} \hspace{-2mm}- \hspace{-1mm}\left(\hspace{-0.3mm}\xi_k\ell_k+\theta_k\hspace{-1mm}\sum_{i\in \mathcal{I}_1}\ell_{i}^2\right)^{-1}\hspace{-4mm}+\epsilon_k,\forall j\in \mathcal{I}_2,
	\end{split}
	\end{equation}
	\vspace{-1mm}
	\begin{equation}
	\label{partial}
	\begin{split}
	&\left(\xi_k\ell_k+\theta_k\sum_{i\in \mathcal{I}_1}\ell_{i}^2\right)^{-1}-\sum_{j\in\mathcal{I}_2\cup\{k\}}d_j^{\max}<\epsilon_k<\left(\xi_k\ell_k+\theta_k\sum_{i\in \mathcal{I}_1}\ell_{i}^2\right)^{-1}-\sum_{j\in\mathcal{I}_2}d_j^{\max},
	\end{split}
	\end{equation}
	where $\mathcal{I}_1\cup\mathcal{I}_2\cup\{k\}=\mathcal{I}$.
	
	\normalsize
	\vspace{-2mm}
	\rule[0.15\baselineskip]{\textwidth}{0.03em}
	\end{figure*} 
	 there exists an equilibrium: 
	\begin{equation}
	\label{opsize}
	\begin{split}
	d_i^*&=0,\forall i\in \mathcal{I}_1,\\
	d_k^*&=\hspace{-1mm}\left(\xi_k\ell_k+\theta_k\sum_{i\in \mathcal{I}_1}\ell_{i}^2\right)^{-1}\hspace{-2mm}-\hspace{-1mm}\sum_{j\in\mathcal{I}_2}\hspace{-1mm}d_j^{\max}\hspace{-1mm}-\epsilon_k  \in\hspace{-1mm}(0,d_k^{\max}),\\
	d_j^*&=d_j^{\max},\forall j\in \mathcal{I}_2.
	\end{split}
	\end{equation}
\end{proposition}

Proposition \ref{different} shows that heterogeneous users may employ different  strategies at equilibrium, including full, partial, and no data revocations. Each user's strategy depends on both its and others' attributes. 
Specifically,  conditions \eqref{nor} and  \eqref{allr} share similar insights with \eqref{no} and \eqref{all},  where users need to trade off model independence indexes, costs, and model performance. For the right-hand side of \eqref{nor} and  \eqref{allr}, the first term represents user $i$'s costs (including privacy cost and unlearning cost), and the remaining two terms reflect the  
model performance. 
Condition \eqref{partial} represents a middle case between  \eqref{nor} and  \eqref{allr} (i.e., user $k$ has medium model independence, costs, and model performance), so user $k$ chooses partial revocation. 
Note that there can be multiple users who partially revoke data (as in Section \ref{strategy}), but we refrain from including the complex conditions when this occurs  for brevity.

We next present a general approach to computing Nash equilibria. 
\subsubsection{General Approach for  Nash Equilibrium Computation}

\begin{figure}[tpb]
	\centering
	\vspace{-6mm}
	\includegraphics[width=2.8 in]{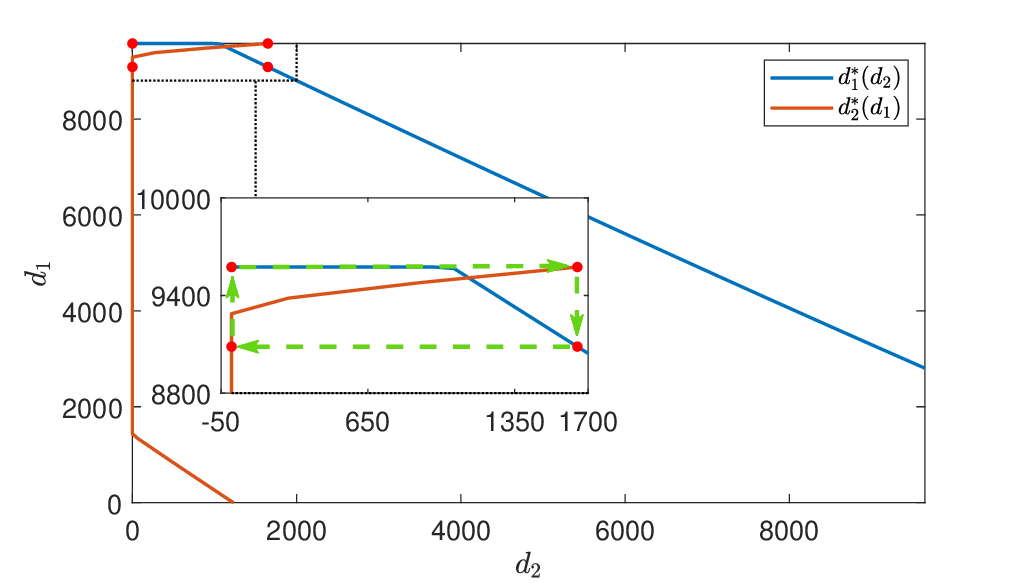}
	\vspace{-2mm}
	\caption{Divergence of sequential best response update: an example of two users. The two users' best responses keep  circulating among the four red points along the green trajectory, unable to converge to a Nash equilibrium where no user has an incentive to alter its strategy.}
 \vspace{-4.5mm}
		\label{d3}
\end{figure}

Many methods in the literature for calculating   Nash equilibria are not applicable to our problem. 
\begin{itemize}
    \item A typical method  involves best response updates simultaneously or sequentially (e.g., \cite{ramazi2017asynchronous}). 
However, such a method  does not necessarily converge for our problem, as shown by the example in Fig.~\ref{d3}. 
\item Another widely-adopted method is gradient (better) response updates (e.g., \cite{lei2022distributed}). However, the identification of appropriate stepsizes under different parameter settings for this problem remains unclear. 
A  small stepsize can cause very slow convergence while   a large stepsize may fail to converge  at all. Moreover, the literature usually requires some technical assumptions for  convergence (such as diagonally strictly concave and asymptotically stable), which do not apply in our context.
\end{itemize}
	
%
%
%

To ensure efficient convergence without requiring any assumption,  we propose a general approach for obtaining the Nash equilibrium next. 
We first define the following function for the convenience of presentation:
\vspace{-1mm}
\begin{equation}
\begin{split}
&W_i(\boldsymbol{d_{-i}})\triangleq \ln\left(d_i^*(\boldsymbol{d_{-i}})+\sum_{j\in\mathcal{I},j\neq i}d_j+\epsilon_i\right)\\&-\xi_id_i^*(\boldsymbol{d_{-i}})\ell_i-\theta_i d_i^*(\boldsymbol{d_{-i}})\sum_{j\in\mathcal{I}, j\neq i}\left(1-\frac{d_j}{d_j^{max}}\right)\ell_j^2.
\end{split}
\end{equation}
This function is obtained by substituting each user $i$'s best response $d_i^*(\boldsymbol{d_{-i}})$ in  \eqref{bere} into its  payoff function $U_i(d_i,\boldsymbol{d_{-i}})$ in \eqref{payofff}.

Inspired by \cite{Chernov2019}, we transform calculating the equilibrium of Game \ref{game1} into solving the following optimization problem:
\vspace{-1mm}
\begin{problem}
	\label{p1}
	\begin{equation*}
	\begin{split}
	\min \;&  \sum_{i\in\mathcal{I}}\left(W_i(\boldsymbol{d_{-i}})-U_i(d_i,\boldsymbol{d_{-i}})\right)\\
	\rm{s.t.}\; 
	&0\le d_i\le d_i^{\max},\; i\in \mathcal{I}\\
	\rm{var.}\; & \{d_{i}\}_{i\in \mathcal{I}}
	\end{split}
	\end{equation*}
\end{problem}
\begin{theorem}
	\label{alne}
	Problem \ref{p1} has a global solution and the corresponding objective	value (i.e., minimum value) is 0. Moreover, each global minimum point of Problem \ref{p1} is a Nash equilibrium of Game \ref{game1}.
\end{theorem}

Consequently, it suffices to describe a method to solve Problem \ref{p1}.
The minimum point can be easily found through numerous global optimization methods   (e.g., \cite{elsakov2011homogeneous,wang2020parallel,piyavskii1972algorithm}) and  optimization toolboxes (e.g., surrogateopt of MATLAB \cite{surrogateopt}). 

There are several advantages of this approach for finding the equilibrium. First, this approach applies without requiring any additional assumptions on the payoff functions or decision variables. Second, it gives simple stopping criteria for iterative algorithms, as we seek to make the objective value equal to zero.  Third, it also enables the use of local optimization methods.
By checking whether a local optimum has an objective value of zero, one can verify if it is also a global optimum. Lastly, instead of tackling $I$ separate problems and searching for a fixed point, we obtain the equilibrium by solving a single optimization problem.


We will show the corresponding numerical results in Section \ref{simulation}. Next, to present more insights, we consider a special scenario with negligible unlearning costs.

\section{Special Scenario: Negligible Unlearning Cost}
\label{negaligible}
In this section, we consider a special scenario where users' unlearning costs are negligible (e.g., when users have abundant or even surplus computation resources). We formally state this in Assumption \ref{a1}. 
\begin{assumption}
	\label{a1}
	Users' unlearning costs are negligible, i.e., 
	\begin{equation}
	\theta_i d_i \hspace{-1.2mm}\sum_{j\in\mathcal{I}, j\neq i}\hspace{-1mm}\left(\hspace{-1mm}1-\frac{d_j}{d_j^{max}}\hspace{-1mm}\right)\hspace{-0.7mm}\ell_j^2 \rightarrow  0, \forall d_i\in [0,d_i^{\max}], \forall  i\in \mathcal{I}.
	\end{equation}
\end{assumption}
In this case, the payoff function of user $i\in \mathcal{I}$ is
\begin{equation}
\label{payoff2}
U_i(d_i,\boldsymbol{d_{-i}})=\ln\left(\sum_{j\in\mathcal{I}}d_j+\epsilon_i\right)-\xi_id_i\ell_i,
\end{equation}
i.e., users only care about the model performance and  privacy concerns when deciding how much data to revoke. 

The results and analysis  in Section \ref{optimal} still hold after setting $\boldsymbol{\theta}=\boldsymbol{0}$. Next, we  present some additional results and insights.

\begin{lemma}
	\label{br2}
	Under Assumption \ref{a1}, the best response of user $i\in \mathcal{I}$  is 
	\begin{equation}
	\label{bere2}
	\begin{split}
	d_i^*(\boldsymbol{d_{-i}})=\left[ \left(\xi_i \ell_i\right)^{-1} - \sum_{j\in\mathcal{I}, j\neq i}d_j - \epsilon_i\right]_0^{d_i^{\max}}.  
	\end{split}
	\end{equation}
\end{lemma}
Lemma \ref{br2} shows that when unlearning costs are negligible, a user's remaining data size $d_i^*$ tends to decrease in  other users' remaining data sizes $\sum_{j\in\mathcal{I}, j\neq i}d_j$. That is, a user will revoke more data if the others revoke less. This is different from the scenario with  considerable unlearning costs where users may follow others' revocation decisions (Corollary \ref{externality}). This is because when unlearning costs are not negligible, users affect each other  through both  the model performance and the unlearning costs. The former increases in users' remaining data sizes while the latter increases in users' revoked data sizes.



In this context, the users' multi-dimensional heterogeneity still poses challenges to computing the equilibrium. To overcome this, we propose to summarize it into a one-dimensional metric, which can greatly streamline the equilibrium calculation process. 
We define $\left(\xi \ell\right)^{-1}-\epsilon$ as the \emph{remaining metric}  and proceed to rank users based on this metric, as per the guidelines set in Assumption \ref{assume}.
\begin{assumption}
	\label{assume}
	Users follow  a strict ascending order of the remaining metric:
 \begin{equation}
 \label{order}
     \left(\xi_1 \ell_1\right)^{-1} -\epsilon_1 <...<  \left(\xi_I \ell_I\right)^{-1} -\epsilon_I
 \end{equation}
 and have the same maximum   data size:
 \begin{equation}
 d_i^{\max}=d^{\max},\forall i\in \mathcal{I}.
 \end{equation}
\end{assumption}
Given Assumption \ref{assume}, we have the following  relationship among the users' data revocation strategies.
\begin{proposition}
	\label{co}
Under Assumptions \ref{a1} and \ref{assume}, users'  remaining data sizes in equilibrium satisfy:
	\begin{equation}
	\label{sequ}
	d_1^*\le d_2^*\le ...\le d_I^*.
	\end{equation}	
\end{proposition}
Proposition \ref{co} shows that a user  possessing a smaller  remaining metric  (i.e., a smaller index  $i$) will revoke more data in equilibrium  (i.e., a smaller $d_i^*$). This is due to its larger privacy cost $\xi$,  data uniqueness $\ell$, and  model independence $\epsilon$. 
 
Based on Proposition \ref{co}, Theorem \ref{neg} shows that there is a unique equilibrium and that the remaining metric $\left(\xi \ell\right)^{-1}-\epsilon$ determines the users' data revocation decisions at this equilibrium.\footnote{Note that as we consider a strict order of users in \eqref{order}, Theorem \ref{neg} shows at most  one  user who partially revokes data at the equilibrium. Otherwise, users with the same remaining metric can simultaneously choose partial revocation in equilibrium.}
\begin{theorem}
	\label{neg}
Under Assumptions \ref{a1} and \ref{assume},	there exists a unique Nash equilibrium and it falls into one of the following two cases: 
	
\noindent 	(i)	If there exists a user $j$ satisfying
	\begin{equation}
	(I-j)d^{\max}\le \left(\xi_j \ell_j\right)^{-1} -\epsilon_j \le (I-j+1)d^{\max},
 	\label{specials}
	\end{equation} 
	then the Nash equilibrium is
	\begin{equation}
	d_i^*=
	\begin{cases}
	0, &\text{if } i<j,\\
	\left(\xi_i \ell_i\right)^{-1}-(I-j)d^{\max} -\epsilon_i,&\text{if } i=j,\\
	d^{\max},&\text{if } i>j.\\
	\end{cases}
	\end{equation}
	
\noindent	(ii)	If no user satisfies \eqref{specials},  there must exist a user $j$ satisfying
	\begin{equation}
	\left(\xi_j \ell_j\right)^{-1} -\epsilon_j< (I-j)d^{\max}< \left(\xi_{j+1} \ell_{j+1}\right)^{-1} -\epsilon_{j+1},
	\end{equation}
	and the Nash equilibrium  is
	\begin{equation}
	d_i^*=
	\begin{cases}
	0, &\text{if } i\le j,\\
	d^{\max},&\text{if } i>j.\\
	\end{cases}
	\end{equation}
\end{theorem}

In the next section, we investigate whether allowing partial revocation is beneficial to users.


\section{Is allowing partial revocation beneficial?}
\label{compareaf}
In this section, we compare users' payoffs and total remaining data size in the two cases of allowing and forbidding partial data revocation.\footnote{For the forbidding partial revocation case, we just need to change each user $i$'s strategy space from continuous actions $[0,d_i^{\max}]$ to  discrete actions $\{0,d_i^{\max}\}$ and conduct a similar analysis as in previous sections.} Specifically, we first focus on the general scenario where users have unlearning costs in Section \ref{general}. Then, we use the special scenario with negligible unlearning costs to further provide insights in Section \ref{special}.


\subsection{General Scenario: Considerable Unlearning Cost}
\label{general}
One might presume that allowing partial revocation will lead to more contributed data. This is because users who might have chosen to fully revoke their data without this option, might now opt for partial revocation instead. 
One might also presume that allowing partial revocation will increase users' payoffs as users have more choices.
However, as shown in Proposition \ref{compare}, the intuition may not hold. 

\begin{proposition}
	\label{compare}
	Compared with forbidding partial revocation, allowing  partial revocation can increase, decrease, or not change the total amount of remaining data and users' payoffs.
\end{proposition}
This counter-intuitive phenomenon arises due to the strategic interactions of the users in the underlying game.  
Specifically,  allowing partial revocation can also make some   users who do not revoke data in the forbidding partial revocation case  revoke some data,  leading to possibly  reduced total remaining data size. 
The resulting potential loss in payoffs from increasing the users' action space is  similar to \emph{Braess's paradox} in transportation networks \cite{braess2005paradox}. 


\subsection{Special Scenario: Negligible Unlearning Cost}
\label{special}


In the following corollary, we give examples of the possible outcomes in Proposition \ref{compare} under the assumption of negligible unlearning costs.  These examples show that users' payoffs and the remaining data sizes  can either increase, decrease, or remain unchanged when partial revocation is allowed. 
\begin{figure*}
	\centering
	\subfigure[no-full.]{
		\begin{minipage}[t]{0.31\linewidth}
			\centering
			\includegraphics[width=2.4 in]{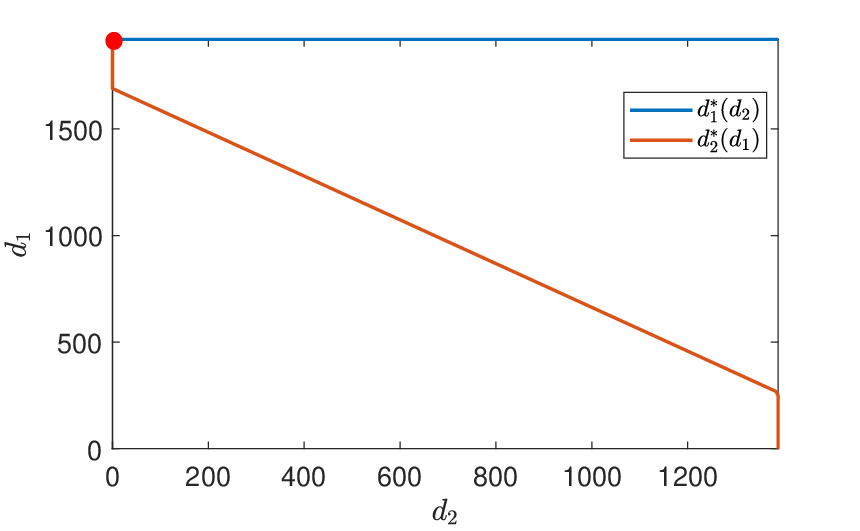}
			\label{0-max}
	\end{minipage}}
 \hspace{0.5mm}
	\subfigure[full-partial.]{
		\begin{minipage}[t]{0.31\linewidth}
			\centering
			\includegraphics[width=2.4 in]{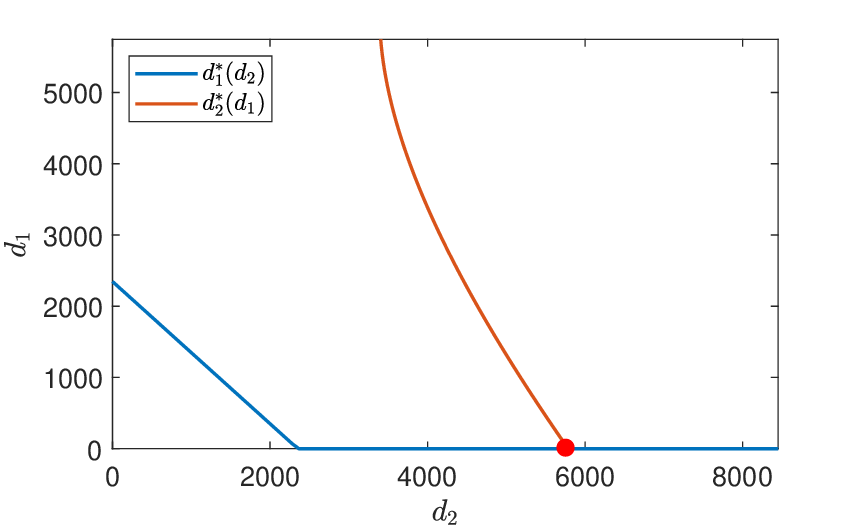}
			\label{p-0}
	\end{minipage}}
	\hspace{0.5mm}
	\subfigure[partial-partial.]{
		\begin{minipage}[t]{0.31\linewidth}
			\centering
			\includegraphics[width=2.4 in]{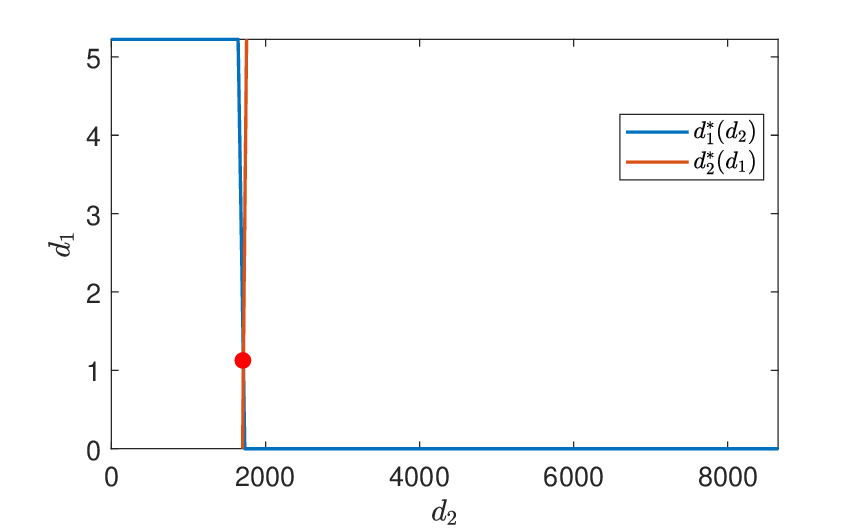}
			\label{p-p}
	\end{minipage}}
 \vspace{-2mm}
	\caption{Three examples of users' best responses to other users'  data revocation strategies when there are two users ($I=2$).  The intersection of the two best response functions (i.e., red dot) in each figure is the Nash equilibrium. The caption of each sub-figure indicates the equilibrium choices of user 1 and user 2, e.g. ``no-full" means that user 1 chooses no revocation while user 2 chooses full revocation.}
	\label{brs}
	\vspace{-5mm}
\end{figure*}

\begin{corollary}
	\label{compare2}
	When users have negligible unlearning costs and are forbidden to partially revoke data,  
	\begin{itemize}
		\item if the following conditions hold: 
		\begin{equation}
		\label{eq1}
		\hspace{-3mm}\ln\hspace{-0.7mm}\frac{(I-j+1)d^{\max}+\epsilon_i}{(I-j)d^{\max}+\epsilon_i}\hspace{-1mm}\ge \hspace{-0.7mm}\xi_i\ell_id^{\max}\hspace{-1mm},\forall i\hspace{-0.7mm}\in \hspace{-1mm}\{j,j+1,...,I\},
		\end{equation}
		user $j$ in Theorem \ref{neg} case (i) will not revoke any data at the equilibrium and other users' equilibrium strategies remain unchanged, i.e.,
		\begin{equation}
		\label{ne1}
		d_{i,forbid}^{*}=
		\begin{cases}
		0, &\text{if } i<j,\\
		d^{\max},&\text{if } i\ge j.\\
		\end{cases}
		\end{equation}
		Moreover, compared to allowing partial revocation, user $j$'s payoff decreases while the other users' payoffs increase, and the total remaining data size also increases.

		\item if the following conditions hold: 
		\begin{equation}
		\label{eq2}
		\hspace{-1mm}\ln\frac{(I-j+1)d^{\max}+\epsilon_i}{(I-j)d^{\max}+\epsilon_i}\hspace{-1mm}\le \xi_i\ell_id^{\max},\forall i\in\hspace{-1mm} \{1,2,...,j\},
		\end{equation}
		user $j$ in Theorem \ref{neg} case (i)  will fully revoke its data at the equilibrium and the other users' equilibrium strategies remain unchanged, i.e.,
		\begin{equation}
		\label{ne2}
		d_{i,forbid}^*=
		\begin{cases}
		0, &\text{if } i\le j,\\
		d^{\max},&\text{if } i> j.\\
		\end{cases}
		\end{equation}
		Compared to allowing partial revocation, all users' payoffs will decrease, and the total remaining data size will also decrease.
		
		\item for Theorem \ref{neg} case (ii), the equilibrium strategies of	users   do not change. Moreover, all users' payoffs and the total remaining data size also remain unchanged.
	\end{itemize}	
\end{corollary}

The condition \eqref{eq1} indicates that users with indexes larger than or equal to $j$ will not deviate from the equilibrium in \eqref{ne1}. The conditions for users with indexes smaller than $j$  are naturally satisfied.  In this case, user $j$' strategy changes from partial revocation (when allowing partial revocation) to no revocation (when forbidding partial revocation),  while the strategies of other users remain constant. This results in   more data  remaining in the system. As a consequence of the improved model performance, other users experience an increase in their payoffs. Nevertheless, due to the constraints imposed on the strategic space, the payoff for user $j$ sees a reduction. 
The insights are similar for the remaining two cases.

\section{Simulations}
\label{simulation}
 In this section, we   perform numerical experiments to   validate our analytical results and reveal more insights.  Specifically, we   first illustrate the best responses of users and  the complexity of computing Nash equilibria in Section \ref{bestr}, then we calculate users'  revocation strategies at the equilibrium in Section \ref{strategy}, and finally we compare the allowing and forbidding partial revocation in Section \ref{simucomp}.

\subsection{Best Response}
\label{bestr}
To illustrate the complexity of calculating Nash equilibria based on users' best responses, we first consider $I=2$ users in the system with randomly generated attribute parameters (i.e., $\ell_i$, $\xi_i$, $\theta_i$, $\epsilon_i$, and $d^{\max}_i$).\footnote{The results and insights  under different groups of randomly generated parameters are similar, so we only present representative  groups of them in this paper. Moreover, unless otherwise specified, we focus  on the general scenario where users have unlearning costs in simulations.} 

Each user may adopt one of three types of strategies at the equilibrium: full revocation, partial revocation, and no revocation, each with different expressions as in Lemma \ref{br}. Thus, even for only two users, we can generally divide the equilibrium calculation into nine categories.
Examples in three of these categories are shown in  Fig.~\ref{brs}.   
Moreover, when their attribute parameters take different values,   users' best response expressions can have different properties, such as non-convex (e.g., Fig.~\ref{0-max} and Fig.~\ref{p-p}), non-linear (e.g., Fig.~\ref{p-0}), or non-monotonic (e.g., Fig.~\ref{d3}),   complicating analysis of the equilibrium. 

\begin{figure*} 
		\begin{minipage}[t]{0.53\linewidth}
			\centering
   \includegraphics[width=2.8 in]{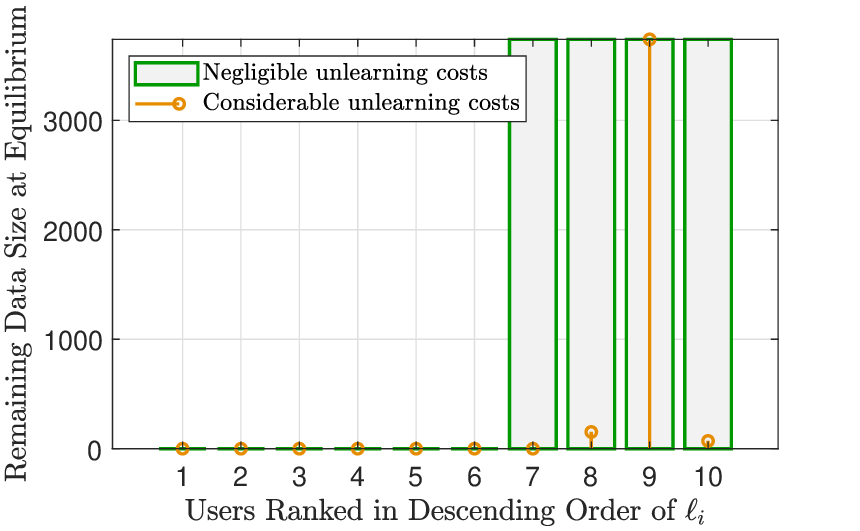}
   \vspace{-2mm}
	\caption{Users' data revocation strategies at the equilibrium   with or without unlearning costs.}
 \vspace{-3mm}
 	\label{f2}
	\end{minipage}
\hspace{1mm}
	\begin{minipage}[t]{0.45\linewidth}
			\centering
\includegraphics[width=2.8 in]{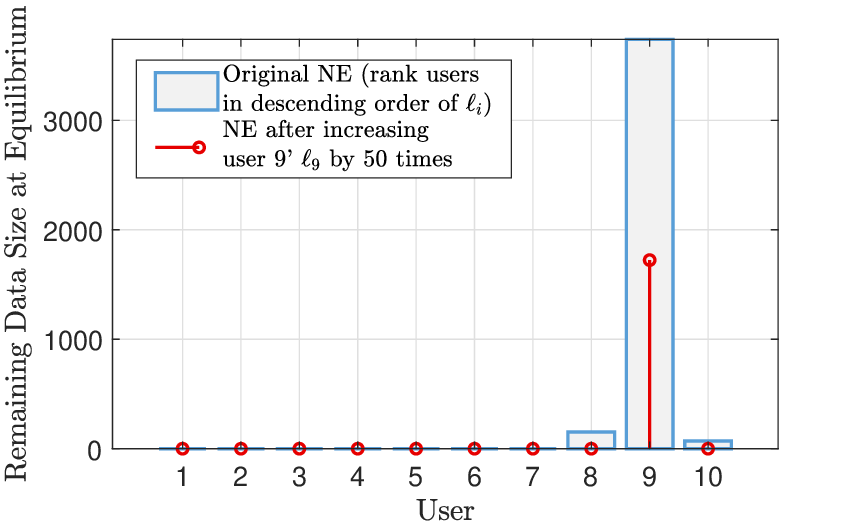}	
   \vspace{-2mm}
   \caption{Users' data  revocation strategies at the Nash equilibrium (NE) before and after   user $9$'s data uniqueness level changes,  when users have unlearning costs.}
   \vspace{-3mm}
			\label{f1}
	\end{minipage}
\end{figure*}
\begin{figure*} 
	\centering
	\subfigure[Difference of total remaining data sizes in two cases (allow$-$forbid)   and difference of users' total payoffs in two cases (allow$-$forbid), when users' data uniqueness levels are multiplied by $k$.]{
		\begin{minipage}[t]{0.52\linewidth}
			\centering
			\includegraphics[width=2.8 in]{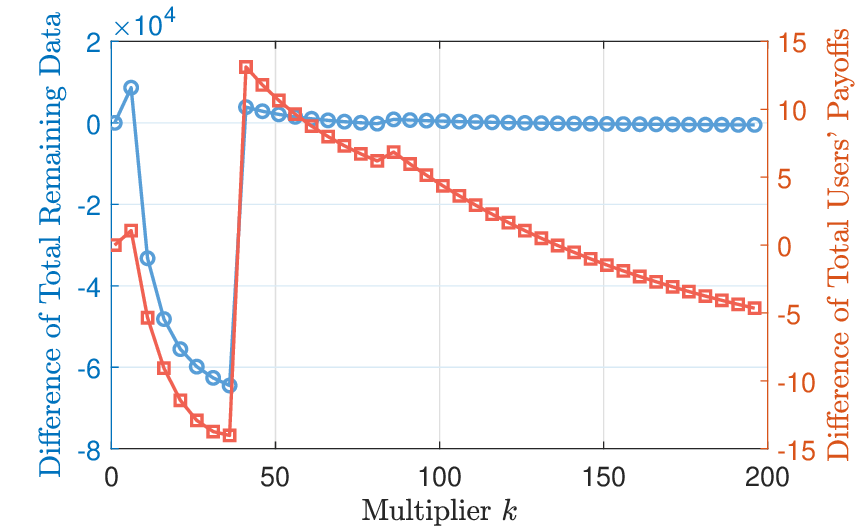}
			\label{d}
	\end{minipage}}
	\hspace{0.5mm}
	\subfigure[Difference of each user's payoffs in two cases (allow$-$forbid) when multiplier $k=6$.]{
		\begin{minipage}[t]{0.45\linewidth}
			\centering
			\includegraphics[width=2.8 in]{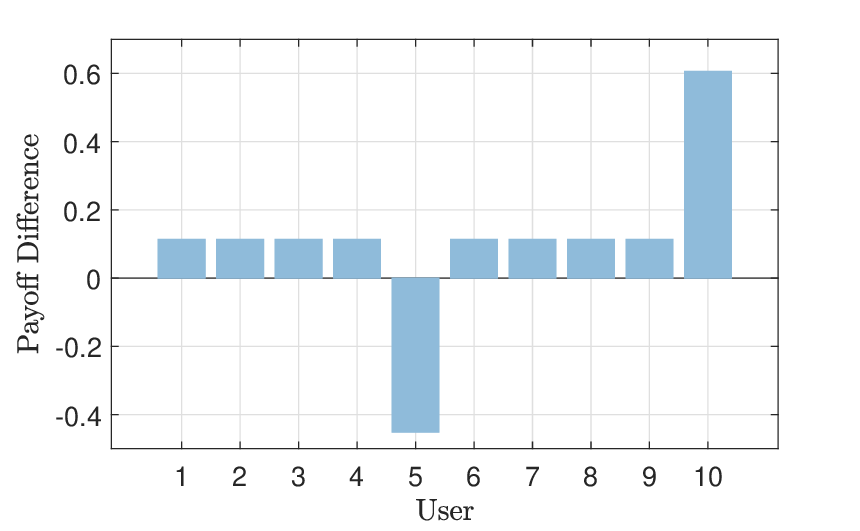}
			\label{d2}
	\end{minipage}}
 \vspace{-2mm}
	\caption{Comparison between allowing and forbidding partial data revocation in terms of users' payoffs and total remaining data size.}
 \vspace{-2mm}
	\label{com}
\end{figure*}
\subsection{Users' Data Revocation Strategies}
\label{strategy}
We use the approach in Theorem \ref{alne} to calculate the equilibria with $I=10$ users.  We randomly generate  users' different data uniqueness levels $\boldsymbol{\ell}$ keeping the other parameters fixed.\footnote{Besides the data uniqueness level, we have similarly explored varying the other parameters.   The results are similar and are omitted due to space limitations. } 

Fig.~\ref{f2} shows users'  data revocation strategies at the equilibrium when they have or do not have unlearning costs. When unlearning costs are negligible (green bar), users with smaller data uniqueness $\ell_i$ (i.e., users 7-10) keep more data in the system, and all users either fully revoke or do not revoke data at equilibrium. This validates the results in Proposition \ref{co} and Theorem \ref{neg}. When users have unlearning costs (orange dot), users' remaining data sizes do  not follow a strict order but roughly show  that users with larger data uniqueness revoke more data, as indicated in Lemma \ref{br}.
The lack of ordering in terms of data sizes is  because, as shown in \eqref{10}, users' data sizes and data uniqueness levels affect each other in a non-monotonic way.

Fig.~\ref{f1} shows how a user's data uniqueness level affects the other users' data revocation strategies when users have unlearning costs. After we increase user 9's data uniqueness level by 50 times, user 9's equilibrium strategy changes from no revocation (blue bar) to partial revocation (redpoint). Users 8 and 10 also revoke more data, i.e., they change their strategies from partial revocation to full revocation. Other users still revoke all their data. This validates the insights of Corollay \ref{externality} that when users have unlearning costs, users may follow  the data revocation decisions of others who have large data uniqueness levels and remaining data sizes.



\subsection{Compare Allowing and Forbidding Partial Revocation}
\label{simucomp}
We numerically compare users' payoffs and remaining data sizes in the two cases of allowing and forbidding partial revocation. Fig.~\ref{d} shows the differences in payoffs and remaining data sizes between the two cases (allowing case minus forbidding case), when users' randomly generated data uniqueness levels are multiplied by $k$. The differences can take positive, zero, or negative values.
This confirms that allowing partial data revocation can either enhance, leave unchanged, or diminish users' total payoff and remaining data size compared with when it is forbidden. 

Fig.~\ref{d2} further shows that 10 users may have different preferences on the two revocation cases. User 5 prefers forbidding partial revocation while others prefer allowing partial revocation. The comparison insights are consistent with Propositions \ref{compare} and Corollary \ref{compare2}.

\section{Conclusion}
\label{conclusion}
This paper focuses on  data revocation in federated unlearning. 
To the best of our knowledge, this is the first study of users' strategic data revocation in federated unlearning with partial revocation allowed. We propose a game theoretic framework to  capture users' decision-making trade-offs among the model performance, data privacy concerns, and federated  unlearning costs. Despite inherent challenges, we  obtain the Nash equilibrium of the game through  both  analytical and numerical methods. We use this equilibrium to deliver some interesting insights, such as that  allowing partial data revocation may not always benefit users compared to when it is not allowed, and users' unlearning costs can cause their cascaded data revocation. 
Possible future work includes the design of incentive mechanisms to improve the users' payoffs. 

\bibliographystyle{IEEEtran}
\bibliography{ref} 

\end{document}